\documentclass[aps,prd,onecolumn,superscriptaddress,nofootinbib]{revtex4-2}
\usepackage[utf8]{inputenc}
\usepackage{amsbsy}
\usepackage{amsmath}
\usepackage{amsfonts}
\usepackage{xcolor}
\usepackage{physics}
\usepackage{graphicx}

\DeclareMathOperator{\sign}{sign}
\begin{document}
\title{Condensed vacuum generated by spin-spin interaction as a source of axial current}

\author{A. Capolupo}
\email{capolupo@sa.infn.it}
\affiliation{Dipartimento di Fisica ``E.R. Caianiello'' Universit\'a degli Studi di Salerno,
  and INFN   Gruppo Collegato di Salerno, Via Giovanni Paolo II, 132, 84084 Fisciano (SA), Italy}

\author{A. Quaranta}
\email{anquaranta@unisa.it}
\affiliation{Dipartimento di Fisica ``E.R. Caianiello'' Universit\'a degli Studi  di Salerno,
  and INFN   Gruppo Collegato di Salerno, Via Giovanni Paolo II, 132, 84084 Fisciano (SA), Italy}
\date{\today}

\begin{abstract}

We reveal the presence of a new source of axial current due to the condensed vacuum generated by  the
spin-spin interaction.
To show this, we consider a quartic Dirac Lagrangian containing a spin-spin interaction term, possibly originating from torsion in Einstein-Cartan-like theories. We use a mean field approach to analyze the quantized theory. We show that the diagonalization of the field Hamiltonian defines a new vacuum state, energetically favored with respect to the free vacuum. Such a vacuum, which is a condensate of particle-antiparticle pairs, is characterized by a nontrivial expectation value of the axial current operator. The new source of axial current, here obtained, can have effects both at the atomic level and at the astrophysical-cosmological level depending on the origin of the spin-spin interaction term.
The  condensate may affect the dark sector of the universe at cosmological level and the axial current  originated by the vacuum condensate could be analyzed in next table top experiments on graphene.

\end{abstract}

\maketitle

\section{Introduction}

Condensation phenomena are ubiquitous in physics. In Quantum Chromodynamics the formation of quark condensates due to the strong interaction is responsible for the spontaneous breakdown of chiral symmetry \cite{QCD,QCD2}. Superfluids \cite{Superfluid,Superfluid2,Superfluid3,Superfluid4,Superfluid5} and superconductors \cite{BCS1,BCS2,BCS3,BCS4} are likewise characterized by a condensed vacuum state. Particle creation phenomena such as the Hawking-Unruh effect \cite{Hawking,Unruh,Takagi,Vanzella}, the Parker effect \cite{Parker,Parker2} as well as the emergence of the Casimir force \cite{Casimir,Casimir2,Casimir3} and the flavor vacuum in particle mixing \cite{Neut1,CapDark1,CapDark2,CapDark3,CapDark4,CapDark5,K1,K2,K3,K4,K5,Cap1,Cap2,Axion4,Neut2,Neut3,Neut4,Neut5,Neut6} can all be understood in terms of a condensed vacuum state.
As it is the case, e. g., in superconductivity, the appearance of condensates can be associated with a phase transition, and thus with the spontaneous breakdown of a symmetry. The vacuum of the theory acquires a non-trivial energy-momentum content in virtue of its condensate structure, leading, in the case of superconductivity, to an energy gap with respect to the ordinary phase.

On the other hand a concept which emerges in differential geometry and applies to several different systems, ranging from
cosmology to condensed matter and particle physics is that of torsion \cite{Torsion1,Hehl1976,Barker1978,Torsion2,Torsion3,Torsion4,Torsion5,Torsion6,Propagating1,Propagating2,Propagating3}.
Many aspects of torsion play crucial roles in  the physics of  materials like
graphene, in the spacetime geometry of the early Universe and in supergravity (SUGRA) theories \cite{SUGRA}. The torsion effects in such contexts are in principle experimentally testable and graphene, where torsion is associated with  appropriate dislocations in the material, gives the opportunity to test  the predictions for the high-energy theories in a controllable laboratory environment.

Torsion is responsible for the appearance of a spin-spin interaction among Dirac fields. The purpose of this work is to analyze the possible formation of fermion condensates due to such kind of spin-spin interactions.
We reveal the presence of a new source of axial current due to the condensed vacuum generated by the
spin-spin interaction.
The theory we discuss is described by the Lagrangian for the Dirac field with an additional pseudovector interaction which is quartic in the fermion field. Such a interaction may emerge also as a consequence of spacetime torsion.
By employing a mean field approach, we effectively transform the quartic field Hamiltonian to a quadratic one, which can be diagonalized upon quantization of the Dirac field. Such diagonalization, achieved through an appropriate Bogoliubov transformation, defines a condensed vacuum, which is unitarily inequivalent to the vacuum of the free theory, in the limit of infinite volume. In particular, the inequivalence between the two representations induces an expectation value of the axial current on the condensed vacuum which is different from zero.
A set of self-consistency equations for the mean field is derived and the energy gap between the free vacuum  and the condensed vacuum is computed.
We further discuss the range of validity of the mean field, showing that it leads to a condition on the value of the ultraviolet cutoff $Q_{UV}$.

The vacuum contribution to the axial current, which we show, represents a source of axial current due to a purely quantum effect, which can manifest itself in different physical phenomena and at different scales, from atomic to astrophysical-cosmological scale, depending on the origin of the spin-spin interaction. For example, it can be due to coupling with an external field or due to torsion in the Einstein-Cartan theory.
Indeed, while the origin of the spin-spin interaction term may be any, it is in particular a byproduct of gravitational theories with torsion.
Next table top experiments on graphene could also allow to study the new source of axial current (and indirectly, of torsion) originated by the vacuum condensate presented in this work.

Considering gravitational theories including a torsion term, as Einstein-Cartan, the condensate may bring along a new contribution to the gravitational field equations, possibly with an impact on the dark sector of the universe. This is apparent, in the present flat spacetime context, from the appearance of an energy gap. Depending on the nature of the underlying torsional theory, possibly featuring a propagating torsion \cite{Propagating1,Propagating2,Propagating3}, the condensate here described may, to different degrees, affect the corresponding cosmological models.

For simplicity, in our treatment we will consider  massless fields, the generalization to massive fields representing a fairly straighforward extension. Moreover,  the theory is set in Minkowski background, while its extension to curved backgrounds shall be pursued in future works. The latter generalization shall certainly modify the details of the computation, nonetheless leaving the essential features of the theory unaltered.

We also note that the spin-spin interaction term is loosely reminiscent of the four-fermion interaction in the BCS theory of superconductivity. We build on such loose analogy pursuing a mean field approach and highlighting, where present, the similiarities. Another close relative of the axial vector interaction is the Heisenberg model of (anti-)ferromagnetism (see e.g \cite{Heis1,Heis2}), since the former effectively reduces to a spin-spin interaction in the non-relativistic limit. In many respects the axial vector interaction does indeed represent the relativistc quantum field theory equivalent of a Heisenberg model. Although considerations regarding the phase transition are not of interest to us in the current work, it is worth noting that the same kind of symmetry (the $SO(3)$ rotational symmetry) is spontaneously broken by the vacuum condensate that emerges due to the axial interaction.

The paper is structured as follows. In section II, we introduce the
quartic fermion Lagrangian and we quantize the Dirac field. Then we diagonalize the Hamiltonian by means of a Bogoliubov transformation and introduce the vacuum condensate. In section III, we derive the vacuum expectation value of axial current on the vacuum condensate and obtain self-consistency equations. The energy gap between the original vacuum and the condensed one, together with considerations on the validity of the mean field approach are also contained in this section. The last section is devoted to the
conclusions. In the appendix A, we report the computations of the angular integrals appearing in the calculus of the vacuum expectation value of the axial current, while some details of the computation of the vacuum expectation value of $J^{\mu 5}J^{ 5}_{\mu}$ are presented in the appendix B.

\section{The Model}

We begin our analysis by considering a Dirac lagrangian $\mathcal{L}$ containing the quartic spin-spin interaction term (possibly due to torsion in Einstein-Cartan theory).
We use the mean field approach to transform the quartic Hamiltonian $\mathcal{H}$ corresponding to $\mathcal{L}$ in a quadratic one. We quantize the Dirac field and diagonalize $\mathcal{H}$, for simplicity, in the case of massless fields.

Our starting point is the quartic Fermion Lagrangian in flat spacetime (the mostly minus signature is employed):
\begin{equation}\label{Lagrangian}
 \mathcal{L} = \mathcal{L}_0 + \mathcal{L}_{Spin-Spin} = \frac{i}{2} \left( \bar{\psi} \gamma^{\mu} \partial_{\mu} \psi - \partial_{\mu} \bar{\psi} \gamma^{\mu} \psi \right) - m\bar{\psi} \psi + \lambda \bar{\psi} \gamma^{5} \gamma^{\mu} \psi \bar{\psi} \gamma^{5} \gamma_{\mu} \psi \ ,
\end{equation}
where $m$ is the mass of the Dirac field, $\gamma^5 = i \gamma^0 \gamma^1 \gamma^2 \gamma^3$ and we use the Dirac representation of the gamma matrices. The coupling constant $\lambda$ depends on the origin of the spin-spin interaction. For example, for the contact interaction induced by torsion in the Einstein-Cartan theory \cite{Hehl1976}, one has $\lambda = - \frac{3}{8} l_P^2$, with $l_P = \sqrt{\frac{\hbar G}{c^3}}$ the Planck length\footnote{Notice the opposite metric signature when comparing to ref. \cite{Hehl1976}}. The sign of $\lambda$ in particular is essential. Setting $S^{\mu} = \frac{1}{2} \bar{\psi} \gamma^5 \gamma^{\mu} \psi$, the spin-spin interaction reads $\mathcal{L}_{Spin-Spin} = 4 \lambda S^{\mu}S_{\mu}$ and the corresponding Hamiltonian is
\begin{equation}\label{Spin-Spin Hamiltonian}
 \mathcal{H}_{Spin-Spin} = -4\lambda S^{\mu}S_{\mu} = -4\lambda (S^0)^2 + 4 \lambda \pmb{S} \cdot \pmb{S}
\end{equation}
with the $3$-vectors denoted in boldface. It is clear from Eq. \eqref{Spin-Spin Hamiltonian} that the interaction tends to align the spins when $\lambda < 0$, see also ref. \cite{Barker1978}. We can intuitively expect that for $\lambda < 0$ this interaction can determine a vacuum state with non-vanishing spin-density, and this will be confirmed later on.

In order to characterize the quantum field theory \eqref{Lagrangian} we employ a mean field approach. Letting $J^{5 \mu} = \bar{\psi} \gamma^{5} \gamma^{\mu} \psi$ the axial current, we assume that the vacuum state develops a non-zero expactation value
\begin{equation}\label{SpinVeV}
 L^{\mu} (t) = \langle J^{5 \mu}\rangle \  = \bra{0_C(t)}J^{5 \mu} \ket{0_C(t)} .
\end{equation}
Here the vacuum state $\ket{0_C(t)}$ is distinct from the free vacuum $\ket{0}$. Its emergence and its properties will be displayed below. Notice that we allow for an arbitrary time dependence, but we assume that the vacuum expectation value (vev) is homogeneous and isotropic. Of course more general spacetime dependencies can be considered, yet leading to a significant complication of the analysis.

The key assumption of the mean field is that deviations from the vev are negligible $\delta J^{5 \mu} = J^{5 \mu} - L^{\mu} (t) \sim 0$.
This allows us to write
 \begin{equation}
 J^{5\mu} J^5_{\mu} = \delta J^{5 \mu} \delta J^{5}_{\mu}+ 2 J^{5\mu} L_{\mu}(t) - L_{\mu}(t) L^{\mu}(t)\simeq 2 J^{5\mu} L_{\mu}(t) - L_{\mu}(t) L^{\mu}(t)
 \end{equation}
 and to make the field Hamiltonian quadratic as
\begin{equation}\label{Hamiltonian}
 \mathcal{H} = \bar{\psi} \left( - i \pmb{\gamma} \cdot \pmb{\nabla} + m \right) \psi - 2 \lambda L_{\mu} (t) \bar{\psi} \gamma^5 \gamma^{\mu} \psi + \lambda L^{\mu}(t) L_{\mu} (t) \ .
\end{equation}
Our task is now to determine the vev $L_{\mu} (t)$ self-consistently. Quantizing the Dirac field and diagonalizing the Hamiltonian \eqref{Hamiltonian} we shall arrive at a set of self-consistency equations, analogous to the gap equations in the BCS theory of superconductivity.

\subsection{The Dirac Equation}

The Dirac equation resulting from the Hamiltonian \eqref{Hamiltonian} is
\begin{equation}\label{DiracEquation}
 i \partial_0 \psi =\left(- i \pmb{\alpha}\cdot \pmb{\nabla} + m \gamma^0 \right) \psi + 2\lambda L_0 (t) \gamma^5 \psi - 2 \lambda \pmb{L} (t) \cdot \pmb{\Sigma} \psi
\end{equation}
where we have introduced the matrix
\begin{equation*}
 \pmb{\Sigma} = \begin{pmatrix}
                 \pmb{\sigma} & 0 \\
                 0 & \pmb{\sigma}
                \end{pmatrix}
\end{equation*}
and $\pmb{\sigma}$ is the vector of Pauli matrices. It is convenient to use the helicity eigenspinors $\xi_s (\hat{p})$, satisfying the basic eigenvalue equation
\begin{equation} \label{spinorProperties}
 (\pmb{\sigma} \cdot \pmb{p}) \xi_s (\hat{p}) = s p \xi_s (\hat{p})
\end{equation}
for $s = \pm 1$, $p = |\pmb{p}|$ and $\hat{p} = \frac{\pmb{p}}{p}$. A specific parametrization of $\xi_s (\hat{p})$ will be used later on, as dictated by convenience. The ansatz for solving eq. \eqref{DiracEquation} is the plane wave
\begin{equation}\label{SolutionAnsatz}
 u_{\pmb{p},s} (t, \pmb{x}) =  e^{i \pmb{p} \cdot \pmb{x}}\begin{pmatrix} f_{\pmb{p},s}(t) \xi_s (\hat{p}) \\ g_{\pmb{p},s} (t) s \xi_s (\hat{p})
                 \end{pmatrix}
\end{equation}
where $f_{\pmb{p},s}, g_{\pmb{p},s}$ are functions of time to be determined. Once the positive energy solution is obtained, the corresponding antiparticle solution is simply given by charge conjugation as
\begin{equation}\label{SolutionAnsatz1}
 v_{\pmb{p},s} (t, \pmb{x}) =  e^{i \pmb{p} \cdot \pmb{x}}\begin{pmatrix} g^*_{\pmb{p},s}(t) \xi_s (\hat{p}) \\ -f^*_{\pmb{p},s} (t) s \xi_s (\hat{p}) \end{pmatrix} \ .
\end{equation}
Inserting the ansatz of Eq. \eqref{SolutionAnsatz} in Eq.\eqref{DiracEquation} and using the defining property of the helicity eigenspinors Eq. \eqref{spinorProperties}, we arrive at the system of equations
\begin{equation}\label{DiracSystem}
 i \begin{pmatrix}
    \partial_t f_{\pmb{p},s} (t) \\ \partial_t g_{\pmb{p},s} (t)
   \end{pmatrix} = \begin{pmatrix} m - 2 \lambda s \pmb{L} (t) \cdot \hat{p} & p + 2 \lambda s L_0 (t) \\ p + 2 \lambda s L_0 (t) & -\left(m + 2 \lambda s \pmb{L}(t) \cdot \hat{p} \right)  \end{pmatrix} \begin{pmatrix} f_{\pmb{p},s} (t) \\ g_{\pmb{p},s} (t) \end{pmatrix} = A_{\pmb{p},s} (t) \begin{pmatrix} f_{\pmb{p},s} (t) \\ g_{\pmb{p},s} (t) \end{pmatrix} \ .
\end{equation}
It turns out that the coefficient matrix $A_{\pmb{p},s} (t)$ commutes at different times if $L_0$ is constant in time, thus allowing for a solution in terms of a simple matrix exponential $e^{- i \int dt' A_{\pmb{p},s}(t')}$. The solution is more involved for $L_0 (t)$ an arbitrary function of time. For simplicity we will stick to the hypothesis of constant $L_0$ and we will verify that it is consistent. The normalized (positive energy) solutions of the system are in this case:
\begin{eqnarray}\label{SystemSolution}
 \nonumber f_{\pmb{p},s} (t) &=& \frac{\omega_{p,s} + m}{(2 \pi)^{\frac{3}{2}} \sqrt{(\omega_{p,s} +m)^2 + (p + 2\lambda s L_0)^2}} e^{- i \omega_{p,s}t} e^{-2i \lambda s \hat{p} \cdot \int_0^t \pmb{L}(t') dt'} \\
 g_{\pmb{p},s} (t) &=& \frac{p + 2\lambda s L_0}{(2 \pi)^{\frac{3}{2}} \sqrt{(\omega_{p,s} +m)^2 + (p + 2\lambda s L_0)^2}} e^{- i \omega_{p,s}t} e^{-2i \lambda s \hat{p} \cdot \int_0^t \pmb{L}(t') dt'},
\end{eqnarray}
where the constant, spin-dependent, frequencies are: $\omega_{p,s} = \sqrt{m^2 + (p + 2 \lambda s L_0)^2}$. Then, the Dirac field is expanded as usual
\begin{equation}\label{FieldExpansion}
 \psi (t, \pmb{x}) = \int d^3 p \left( b_{\pmb{p},s} u_{\pmb{p},s} (t, \pmb{x}) + d^{\dagger}_{\pmb{p},s} v_{\pmb{p},s}(t,\pmb{x}) \right)
\end{equation}
with $u_{\pmb{p},s}$ and  $v_{\pmb{p},s}$ given by eqs.\eqref{SolutionAnsatz} and \eqref{SolutionAnsatz1} and  the coefficients satisfying the canonical anticommutation relations.

\subsection{Diagonalizing the field Hamiltonian}

We can now plug in the expansion \eqref{FieldExpansion} in the field Hamiltonian
\eqref{Hamiltonian}. After a lengthy but straightforward computation we can write
\begin{eqnarray}\label{ExpandedFieldHamiltonian}
 \nonumber H &=& \int d^3 x \mathcal{H} = \int d^3 k \Bigg \lbrace \sum_{s} \left[P_{\pmb{k},s}(t) b^{\dagger}_{\pmb{k},s}b_{\pmb{k},s} + Q^{\dagger}_{\pmb{k},s}(t) d_{\pmb{k},s}d^{\dagger}_{\pmb{k},s} + R_{\pmb{k},s}(t) d_{\pmb{k},s}b_{\pmb{k},s} + R^*_{\pmb{k},s}(t) b^{\dagger}_{\pmb{k},s} d^{\dagger}_{\pmb{k},s}  \right] + S_{\pmb{k}}(t) b^{\dagger}_{\pmb{k},+} b_{\pmb{k},-} \\ \nonumber &+& S^*_{\pmb{k}}(t)b^{\dagger}_{\pmb{k},-} b_{\pmb{k},+}
 + V_{\pmb{k}}(t) d_{\pmb{k},-}b_{\pmb{k},+} + V^{*}_{\pmb{k}}(t) b^{\dagger}_{\pmb{k},+}d^{\dagger}_{\pmb{k},-} + U_{\pmb{k}}(t) d_{\pmb{k},+}b_{\pmb{k},-} + U^{*}_{\pmb{k},t}b^{\dagger}_{\pmb{k},-}d^{\dagger}_{\pmb{k},+} + T_{\pmb{k}}(t) d_{\pmb{k},-}d^{\dagger}_{\pmb{k},+} + T^*_{\pmb{k}}(t) d_{\pmb{k},+} d^{\dagger}_{\pmb{k},-} \Bigg \rbrace \\
 &&\end{eqnarray}
where the coefficients $P_{\pmb{k},s} (t)$, $Q_{\pmb{k},s} (t)$, $R_{\pmb{k},s} (t)$, $S_{\pmb{k},s} (t)$, $T_{\pmb{k},s} (t)$, $U_{\pmb{k},s} (t)$,
$V_{\pmb{k},s} (t)$ are given by
\begin{eqnarray}\label{coeff}
 P_{\pmb{k},s} (t) &=& m + \frac{2(k+2\lambda s L_0)(k\omega_{k,s}-2\lambda s m L_0)}{(\omega_{k,s} + m)^2 + (k + 2\lambda s L_0)^2} - 2s\lambda \left \lbrace \hat{k} \cdot \pmb{L}(t) - (2\pi)^3 L_0 \left[ f^*_{\pmb{k},s} (t) g_{\pmb{k},s}(t) + g^*_{\pmb{k},s} (t) f_{\pmb{k},s}(t)  \right] \right \rbrace \\
 Q_{\pmb{k},s} (t) &=& -m - \frac{2(k+2\lambda s L_0)(k\omega_{k,s}-2\lambda s m L_0)}{(\omega_{k,s} + m)^2 + (k + 2\lambda s L_0)^2} - 2s\lambda \left \lbrace \hat{k} \cdot \pmb{L}(t) - (2\pi)^3 L_0 \left[ f^*_{\pmb{k},s} (t) g_{\pmb{k},s}(t) + g^*_{\pmb{k},s} (t) f_{\pmb{k},s}(t)  \right] \right \rbrace \\
 R_{\pmb{k},s} (t) &=& \frac{4m \lambda s L_0 (\omega_{k,s} + m)e^{-2i \omega_{k,s}t}e^{-4i\lambda s \hat{k} \cdot \int_0^t \pmb{L}(t')dt'}}{(\omega_{k,s} + m)^2 + (k + 2\lambda s L_0)^2} - 2s (2\pi)^3 \lambda L_0 \left(f_{\pmb{k},s}^2(t)-g_{\pmb{k},s}^2(t) \right) \\
 S_{\pmb{k}} (t) &=& - 2 \lambda (2\pi)^3 \left(\pmb{\epsilon}(\hat{k}) \cdot \pmb{L}(t) \right)\left(f^{*}_{\pmb{k},+}(t) f_{\pmb{k},-}(t) - g^{*}_{\pmb{k},+}(t) g_{k,-}(t)\right) \\
 T_{\pmb{k}} (t) &=& 2 \lambda (2\pi)^3 \left(\pmb{\epsilon}^*(\hat{k}) \cdot \pmb{L}(t) \right)\left(f^{*}_{\pmb{k},+}(t) f_{\pmb{k},-}(t) - g^{*}_{\pmb{k},+}(t) g_{k,-}(t)\right) \\
 U_{\pmb{k}} (t) &=& - 2 \lambda (2\pi)^3 \left(\pmb{\epsilon}(\hat{k}) \cdot \pmb{L}(t) \right)\left(g_{\pmb{k},+}(t) f_{\pmb{k},-}(t) + f_{\pmb{k},+}(t) g_{k,-}(t)\right) \\
 V_{\pmb{k}} (t) &=& - 2 \lambda (2\pi)^3 \left(\pmb{\epsilon}^*(\hat{k}) \cdot \pmb{L}(t) \right)\left(g_{\pmb{k},+}(t) f_{\pmb{k},-}(t) + f_{\pmb{k},+}(t) g_{k,-}(t)\right).
\end{eqnarray}

Here $f_{\pmb{k},s}, g_{\pmb{k},s}$ are provided by eq. \eqref{SystemSolution} and we have further defined the vector $\pmb{\epsilon}(\hat{k}) = \xi^{\dagger}_{+}(\hat{k}) \pmb{\sigma} \xi_{-} (\hat{k})$.
The Hamiltonian of eq. \eqref{ExpandedFieldHamiltonian} is not diagonal in the field operators and can be diagonalized by means of a proper Bogoliubov transformation. While the latter may be constructed in the general case, its determination is quite involved. A significant simplification occurs in the massless limit $m = 0$, where the mixing of the $4$ operators $b_{\pmb{k},\pm}, d_{\pmb{k}, \pm}$ is reduced to the mixing of only $2$ of them.
In the following, for sake of simplicity, we consider such a case. For $m=0$ the Hamiltonian coefficients \eqref{coeff} become
\begin{eqnarray}\label{MasslessHamiltonianCoefficients}
   && P_{\pmb{k},s} = |k + 2 s \lambda L_0| - 2 s \lambda \pmb{L} (t) \cdot \hat{k}    \ ; \\
   && Q_{\pmb{k},s} = \sign (k + 2 s \lambda L_0) (2s\lambda L_0 -k) - 2s\lambda \pmb{L} (t) \cdot \hat{k} \ ; \\
   && R_{\pmb{k},s} = 0 \ ; \\
  && S_{\pmb{k}} = -  \lambda \left(\pmb{\epsilon}(\hat{k}) \cdot \pmb{L}(t) \right) \left( 1 - \sign (k+2\lambda L_0) \sign(k-2\lambda L_0) \right) e^{i \left(\omega_{k,+} - \omega_{k,-} \right)t}e^{-4i\lambda \hat{k} \cdot \int_0^t \pmb{L}(t') dt'} \ ; \\
 &&   T_{\pmb{k}} = \left( \frac{-\pmb{\epsilon}^*(\hat{k}) \cdot \pmb{L}(t)}{\pmb{\epsilon}(\hat{k}) \cdot \pmb{L}(t) }\right) S_{\pmb{k}} \ ; \\
 && U_{\pmb{k}} =  -  \lambda \left(\pmb{\epsilon}(\hat{k}) \cdot \pmb{L}(t) \right) \left(\sign (k-2\lambda L_0) + \sign(k+ 2\lambda L_0) \right)e^{i \left( \omega_{k,+} + \omega_{k,-}\right)t} \ ;
 \\
 && V_{\pmb{k}} = \left( \frac{\pmb{\epsilon}^*(\hat{k}) \cdot \pmb{L}(t)}{\pmb{\epsilon}(\hat{k}) \cdot \pmb{L}(t) }\right) U_{\pmb{k}}  \ ,
\end{eqnarray}
respectively.
It is further convenient to write the Hamiltonian as $H = \int d^3 k  \mathcal{B}^{\dagger}_{\pmb{k}}\mathcal{H}_{\pmb{k}} \mathcal{B}_{\pmb{k}}$, where $\mathcal{H}_{k}$ is a momentum-dependent hermitian matrix and we define the column vector $\mathcal{B}_{\pmb{k}} = \left( b_{\pmb{k},+}, b_{\pmb{k},-}, d^{\dagger}_{\pmb{k},+}, d^{\dagger}_{\pmb{k},-} \right)^T$. Now, depending on the sign of $\lambda L_0$, we can distinguish three cases. We consider explicitly $\lambda L_0 \geq 0$, but the opposite sign can be treated analogously:
\begin{enumerate}
 \item $k < 2 \lambda L_0$.
In this case $U_{\pmb{k}} = 0 = V_{\pmb{k}}$ and the Hamiltonian matrix $\mathcal{H}_{\pmb{k}}$ is (setting $\mu_{\pmb{k}} = \left(\pmb{\epsilon}(\hat{k}) \cdot \pmb{L}(t) \right)$ )
\begin{equation*}
  \begin{pmatrix}
                          \omega_{k,+}-2\lambda\pmb{L} \cdot \hat{k} & - 2 \lambda \mu_{\pmb{k}} e^{2 i k t} e^{-4i\lambda \hat{k}\cdot \int_0^t \pmb{L}(t') dt'} & 0 & 0 \\ - 2 \lambda \mu^*_{\pmb{k}} e^{-2 i k t} e^{4i\lambda \hat{k}\cdot \int_0^t \pmb{L}(t') dt'} & \omega_{k,-} + 2 \lambda  \pmb{L}(t) \cdot \hat{k} & 0 & 0 \\ 0 & 0 & \omega_{k,-} - 2 \lambda \pmb{L}(t) \hat{k} & 2 \lambda \mu_{\pmb{k}} e^{-2 i k t} e^{4i\lambda \hat{k}\cdot \int_0^t \pmb{L}(t') dt'} \\ 0 & 0 & 2 \lambda \mu^*_{\pmb{k}} e^{2 i k t} e^{-4i\lambda \hat{k}\cdot \int_0^t \pmb{L}(t') dt'} & - \omega_{k,+} + 2 \lambda \pmb{L}(t) \cdot \hat{k}
                         \end{pmatrix}
\end{equation*}
In order to diagonalize this matrix it is sufficient to consider a Bogoliubov transformation that mixes $ b_{\pmb{k},+}$ with  $b_{\pmb{k},-}$ and $d_{\pmb{k},+}$ with $d_{\pmb{k},-}$. Since no creation operator is involved, we can anticipate that these modes (with $k< 2\lambda L_0$) will not contribute to the vev, and the details of the transformation are irrelevant to the final result.

\item $k = 2\lambda L_0$. In this case the Hamiltonian is already diagonal and $\mathcal{H}_{\pmb{k}}$ reads
\begin{equation*}
\begin{pmatrix}
 4 \lambda L_0 - 2 \lambda \pmb{L}(t) \cdot \hat{k} & 0 & 0 & 0 \\ 0 & 2 \lambda \pmb{L}(t) \cdot \hat{k} & 0 & 0 \\ 0 & 0 & - 2 \lambda \pmb{L}(t) \cdot \hat{k} & 0 \\ 0 & 0 & 0 & 2 \lambda \pmb{L}(t) \cdot \hat{k}
\end{pmatrix}
\end{equation*}

\item $k > 2 \lambda L_0$. The Hamiltonian matrix is

\begin{equation}\label{ThirdHamiltonian}
 \begin{pmatrix}
  \omega_{k,+} - 2 \lambda \pmb{L}(t) \cdot \hat{k} & 0 & 0 & - 2 \lambda \mu_{\pmb{k}} e^{2ikt} \\ 0 & \omega_{k,-} + 2 \lambda \pmb{L}(t) \cdot \hat{k} & - 2 \lambda \mu^*_{\pmb{k}} e^{2ikt} & 0 \\ 0 & - 2 \lambda \mu_{\pmb{k}} e^{-2ikt} & 2\lambda L_0 - k - 2 \lambda \pmb{L}(t) \cdot \hat{k} & 0 \\ - 2 \lambda \mu^*_{\pmb{k}} e^{-2ikt} & 0 & 0 & -k - 2\lambda L_0 + 2 \lambda \pmb{L}(t) \cdot \hat{k}
 \end{pmatrix} \ .
\end{equation}
\end{enumerate}
The third case \eqref{ThirdHamiltonian} is the only relevant one to the computation of the vev. For the modes with $k>2\lambda L_0$ the Hamiltonian mixes $b_{\pmb{k},+}$ with $d^{\dagger}_{\pmb{k},-}$ and $b_{\pmb{k},-}$ with $d^{\dagger}_{\pmb{k},+}$. To diagonalize eq. \eqref{ThirdHamiltonian} we postulate a Bogoliubov transformation of the form
\begin{eqnarray}\label{BogoliubovTransformation}
 \nonumber B_{\pmb{k},s} (t) &=& \cos \Theta_{\pmb{k}, s} (t) \ b_{\pmb{k},s} - e^{i \delta_{\pmb{k},s} (t)} \sin \Theta_{\pmb{k},s}(t) \ d^{\dagger}_{\pmb{k},-s} \\ D^{\dagger}_{\pmb{k},s} (t) &=& \cos \Theta_{\pmb{k},s} (t) \ d^{\dagger}_{\pmb{k},s} + e^{-i \delta_{\pmb{k},s} (t)} \sin \Theta_{\pmb{k},s}(t) \ b_{\pmb{k},-s} \ .
\end{eqnarray}
While eqs. \eqref{BogoliubovTransformation} define a canonical transformation by construction, the explicit form of $\Theta_{\pmb{k},s}(t)$ and $\delta_{\pmb{k},s} (t)$ has to be extracted from the requirement of diagonalization, so that in terms of the new operators the Hamiltonian is simply $H = \sum_s \int d^{3}k \left(\mathcal{E}_{\pmb{k},s} (t) B^{\dagger}_{\pmb{k},s} (t) B_{\pmb{k},s}(t) +\tilde{\mathcal{E}}_{\pmb{k},s} (t) D^{\dagger}_{\pmb{k},s} (t) D_{\pmb{k},s}(t)\right)$ for some real $\mathcal{E}_{\pmb{k},s} (t), \tilde{\mathcal{E}}_{\pmb{k},s} (t) $. It is straightforward to check that this is achieved with
\begin{eqnarray}\label{BogoliubovCoefficients}
 \nonumber \sin \Theta_{\pmb{k},s}(t) &=& \frac{4 \lambda |\mu_{\pmb{k}}|}{\sqrt{\left[ 2\left(k+2s\lambda(L_0 - \pmb{L}(t) \cdot \hat{k})\right) + \sqrt{4\left[ k + 2 s \lambda(L_0 - \pmb{L}(t) \cdot \hat{k}) \right]^2 + 16 \lambda^2 |\mu_{\pmb{k}}|^2} \right]^2 + 16 \lambda^2 |\mu_{\pmb{k}}|^2}} \\ e^{i \delta_{\pmb{k},+} (t)} &=& \frac{\mu_{\pmb{k}}}{|\mu_{\pmb{k}}|}e^{2ikt} \  ; \ \ \ \ \ \ \ e^{i \delta_{\pmb{k},-} (t)} = \frac{\mu^*_{\pmb{k}}}{|\mu_{\pmb{k}}|}e^{2ikt} \ ,
\end{eqnarray}
where we recall $\mu_{\pmb{k}} = \left(\pmb{\epsilon}(\hat{k}) \cdot \pmb{L}(t) \right)$. It is important to notice that the Bogoliubov transformation of eq. \eqref{BogoliubovTransformation} does not mix different momentum components and that it does only make sense for the modes with $k>2\lambda L_0$. A distinct transformation holds for the modes with $k< 2\lambda L_0$, yet, as already precised, its details are irrelevant to the computation of the spin vev. The new operators for $k < 2\lambda L_0$, mixing only annihilation operators with distinct helicities, do indeed share the same vacuum as the original operators $b_{\pmb{k},\pm}, d_{\pmb{k},\pm}$, which we denote by $\ket{0}$:
\begin{equation*}
 b_{\pmb{k},\pm} \ket{0} = 0 \ ; \ \ \ \ d_{\pmb{k},\pm} \ket{0} = 0 \ .
\end{equation*}
On the other hand, the Bogoliubov transformation of eq. \eqref{BogoliubovTransformation} defines a new time-dependent vacuum state $\ket{0_C (t)}$ given by
\begin{equation}\label{CondensedVacuum}
B_{\pmb{k},s} (t) \ket{0_C (t)} = 0 \ ; \ \ \ \ D_{\pmb{k},s} (t) \ket{0_C (t)} = 0 \ .
\end{equation}
By the usual arguments \cite{Umezawa1,Umezawa2}, it can be seen that the two vacua belong to unitarily inequivalent representations, and that, in particular, the $\ket{0_C(t)}$ vacuum is a condensate of fermion pairs, with condensation densities proportional to $\sin^2 \Theta_{\pmb{k},s}(t)$ for the two helicities. For further manipulations it is also convenient to define a generator of Bogoliubov transformations $\mathcal{R}(t)$, implicitly given by
\begin{equation}\label{BogoliubovGenerator}
\mathcal{R}^{-1}(t) b_{\pmb{k},s} \mathcal{R}(t) = B_{\pmb{k},s} (t) \ ; \ \ \mathcal{R}^{-1}(t) d_{\pmb{k},s} \mathcal{R}(t) = D_{\pmb{k},s} (t) \ ; \ \ \ket{0_C(t)} = \mathcal{R}^{-1}(t)\ket{0} \ .
\end{equation}
The generator can be determined explicitly by means of the Baker-Campbell-Hausdorff formula \cite{Campbell}, but for our purposes it is sufficient to notice that its general form will be $\mathcal{R}(t) = e^{\Theta_{\pmb{k},s}(t) F(b_{\pmb{k},s}, d_{\pmb{k},s})}$ for some function $F$ of the annihilation operators, as appropriate for a rotation by an angle $\Theta_{\pmb{k},s}(t)$. Its inverse does simply perform the opposite rotation by $-\Theta_{\pmb{k},s}(t)$, so that
\begin{eqnarray}\label{InverseBogoliubov}
 \nonumber\mathcal{R}(t) b_{\pmb{k},s} \mathcal{R}^{-1}(t) &=& \cos \Theta_{\pmb{k},s} (t) \ b_{\pmb{k},s} + e^{i \delta_{\pmb{k},s} (t)} \sin \Theta_{\pmb{k},s}(t) \ d^{\dagger}_{\pmb{k},-s} \\ {R}(t) d^{\dagger}_{\pmb{k},s} \mathcal{R}^{-1}(t)  &=& \cos \Theta_{\pmb{k},s} (t) \ d^{\dagger}_{\pmb{k},s} - e^{-i \delta_{\pmb{k},s} (t)} \sin \Theta_{\pmb{k},s}(t) \ b_{\pmb{k},-s} \ ,
\end{eqnarray}
compare with eqs. \eqref{BogoliubovGenerator} and \eqref{BogoliubovTransformation}.

\section{Vev of the axial current and self-consistency equations}

Using the results obtained above,
we can now give a more precise meaning to Eq. \eqref{SpinVeV} as the expectation value of the axial current on the condensed vacuum $\ket{0_C(t)}$, as defined by diagonalization of the field Hamiltonian:
\begin{equation}\label{SpinVeV2}
L^{\mu}(t) = \bra{0_C(t)} \bar{\psi} \gamma^5 \gamma^\mu \psi \ket{0_C(t)} \ .
\end{equation}
Then, we derive a set of self-consistency equations and compute the energy gap between the free vacuum $\ket{0 }$ and the vacuum $\ket{0_C(t)}$. Moreover, we show that the mean field approach imposes    a condition on the value of the ultraviolet cutoff $Q_{UV}$.

We begin with the $0$ component of eq. \eqref{SpinVeV2}. After plugging the field expansion \eqref{FieldExpansion} we obtain
\begin{eqnarray}\label{ZeroComponent}
 \nonumber  L^0 &=& - \sum_{r,s} \int d^3 k \int d^3 q \ e^{i \left(\pmb{q}-\pmb{k}\right) \cdot \pmb{x}} \Bigg \lbrace A_{\pmb{k},\pmb{q},r,s} \bra{0_C(t)} b^{\dagger}_{\pmb{k},r} b_{\pmb{q},s} \ket{0_C (t)} +  B_{\pmb{k},\pmb{q},r,s} \bra{0_C(t)} b^{\dagger}_{\pmb{k},r} d^{\dagger}_{\pmb{q},s} \ket{0_C (t)} \\
 &+&  C_{\pmb{k},\pmb{q},r,s} \bra{0_C(t)} d_{\pmb{k},r} b_{\pmb{q},s} \ket{0_C (t)} +  D_{\pmb{k},\pmb{q},r,s} \bra{0_C(t)} d_{\pmb{k},r} d^{\dagger}_{\pmb{q},s} \ket{0_C (t)} \Bigg \rbrace \ ,
\end{eqnarray}
where the coefficients are
\begin{eqnarray*}
 A_{\pmb{k},\pmb{q},r,s} &=& \xi^{\dagger}_r (\hat{k}) \xi_s (\hat{q}) \left(sf^*_{\pmb{k},r}g_{\pmb{q},s} + r g^*_{\pmb{k},r} f_{\pmb{q},s}  \right)\ ;
 \\
    B_{\pmb{k},\pmb{q},r,s} &=& \xi^{\dagger}_r (\hat{k}) \xi_s (\hat{q}) \left(-sf^*_{\pmb{k},r}f^*_{\pmb{q},s} + r g^*_{\pmb{k},r} g^*_{\pmb{q},s}  \right) \ ;
     \\
     C_{\pmb{k},\pmb{q},r,s} &=& \xi^{\dagger}_r (\hat{k}) \xi_s (\hat{q}) \left(sg_{\pmb{k},r}g_{\pmb{q},s} - r f_{\pmb{k},r} f_{\pmb{q},s}  \right)\ ; \\
      D_{\pmb{k},\pmb{q},r,s} &=& \xi^{\dagger}_r (\hat{k}) \xi_s (\hat{q}) \left(-sg_{\pmb{k},r}f^*_{\pmb{q},s} - r f_{\pmb{k},r} g^*_{\pmb{q},s}  \right).
\end{eqnarray*}
The expectation values can be computed in a simple fashion using eqs \eqref{BogoliubovGenerator} and \eqref{InverseBogoliubov}:
\begin{eqnarray}\label{SubVev1}
\nonumber && \bra{0_C(t)} b^{\dagger}_{\pmb{k},r} b_{\pmb{q},s} \ket{0_C (t)} = \bra{0} \mathcal{R}(t) b_{\pmb{k},r} b_{\pmb{q},s} \mathcal{R}^{-1}(t) \ket{0} = \bra{0} \mathcal{R}(t) b_{\pmb{k},r}\mathcal{R}^{-1}(t) \mathcal{R}(t) b_{\pmb{q},s} \mathcal{R}^{-1}(t) \ket{0} = \\ \nonumber
&& \bra{0}\left(\cos \Theta_{\pmb{k},r} (t) \ b^{\dagger}_{\pmb{k},r} + e^{-i \delta_{\pmb{k},r} (t)} \sin \Theta_r (t) \ d_{\pmb{k},-r} \right) \left(\cos \Theta_{\pmb{q},s} (t) \ b_{\pmb{q},s} + e^{i \delta_{\pmb{q},s} (t)} \sin \Theta_{\pmb{k},s}(t) \ d^{\dagger}_{\pmb{q},-s} \right) \ket{0} \\ && = \delta^3 (\pmb{k}-\pmb{q}) \delta_{rs} \sin^2 \Theta_{\pmb{k},r}
\end{eqnarray}
and similarly
\begin{eqnarray}\label{SubVev2}
\nonumber  \bra{0_C(t)} b^{\dagger}_{\pmb{k},r} d^{\dagger}_{\pmb{q},s} \ket{0_C (t)} &=& \delta^3 (\pmb{k}-\pmb{q}) \delta_{r,-s} \sin \Theta_{\pmb{k},r} \cos \Theta_{\pmb{k},r}e^{-i\delta_{\pmb{k},r}} \\
\nonumber  \bra{0_C(t)} d_{\pmb{k},r} b_{\pmb{q},s} \ket{0_C (t)} &=& \delta^3 (\pmb{k}-\pmb{q}) \delta_{r,-s} \sin \Theta_{\pmb{k},-r} \cos \Theta_{\pmb{k},-r}e^{i\delta_{\pmb{k},-r}} \\
 \bra{0_C(t)} d_{\pmb{k},r} d^{\dagger}_{\pmb{q},s} \ket{0_C (t)}&=& \delta^3 (\pmb{k}-\pmb{q}) \delta_{rs} \cos^2 \Theta_{\pmb{k},-r} \ .
\end{eqnarray}
Inserting eqs. \eqref{SubVev1} and \eqref{SubVev2} in \eqref{ZeroComponent} we obtain
\begin{equation*}
 L^0 = - \sum_r \int d^3 k \left \lbrace A_{\pmb{k},\pmb{k},r,r} \sin^2 \Theta_{\pmb{k},r} + D_{\pmb{k},\pmb{k},r,r} \sin^2 \Theta_{\pmb{k},-r} + \left(B_{\pmb{k},\pmb{k},r,-r} \cos \Theta_{\pmb{k},r} \sin \Theta_{\pmb{k},r} e^{-i\delta_{\pmb{k},r}}  + c.c. \right) \right \rbrace .
\end{equation*}
Considering that $A_{\pmb{k},\pmb{k},r,r}=-D_{\pmb{k},\pmb{k},r,r} = \frac{r}{(2\pi)^3}$ and $B_{\pmb{k},\pmb{k},r,-r}=0=C_{\pmb{k},\pmb{k},r,-r}$, $L^0$ further simplifies to
\begin{equation}\label{FirstSelfConsistency}
 L^0 = -\sum_r \frac{r}{2(\pi)^3} \int d^3 k \left(\cos^2 \Theta_{\pmb{k},r} + \cos^2 \Theta_{\pmb{k},-r} -1   \right) = 0 \ ,
\end{equation}
where the last equality follows noting that the summand is odd in $r$.
Eq.\eqref{FirstSelfConsistency} represents the first self-consistency equation.

Notice that, a priori, the momentum integrals in eqs. \eqref{ZeroComponent} and \eqref{FirstSelfConsistency} extend from $k=2\lambda L_0$ to $\infty$ (since the modes with $k< 2 \lambda L_0$ yield zero contribution). The helicity sum in eq. \eqref{FirstSelfConsistency} vanishes regardless, setting the first self-consistency equation  to the condition $L^0 = 0$. This is consistent with the initial assumption of constant $L^0$. In addition, since $2\lambda L^0 = 0$, it is always the case that $k \geq 2\lambda L^0$, and the momentum integrals for the remaining components must extend in the whole range of momenta.
Following similar steps we can write down the other self-consistency equation as
\begin{eqnarray}\label{SecondSelfConsistency}
\nonumber \pmb{L}(t) &=& - \sum_{r}\frac{r}{(2\pi)^3} \int d^3 k \hat{k} \left(\cos^2 \Theta_{\pmb{k},r} - \cos^2 \Theta_{\pmb{k},-r} +1 \right) \\ &-& \int d^3 k \Bigg \lbrace \frac{\pmb{\epsilon}(\hat{k})e^{2ikt}}{(2\pi)^3} \cos \Theta_{\pmb{k},+} \sin \Theta_{\pmb{k},+} e^{- i \delta_{\pmb{k},+}} +\frac{\pmb{\epsilon}^*(\hat{k})e^{2ikt}}{(2\pi)^3} \cos \Theta_{\pmb{k},-} \sin \Theta_{\pmb{k},-} e^{- i \delta_{\pmb{k},-}} + c. c. \Bigg \rbrace \ .
\end{eqnarray}
Up to now all the computations performed have been non-perturbative. To proceed with the explicit analytical evaluation of the integrals in eq. \eqref{SecondSelfConsistency}, we now perform a perturbative expansion of the integrand with respect to the coupling $\lambda$, truncating at order three. It's easy to check that had we to stop at the first order in $\lambda$ (the second order vanishes identically) only the trivial solution $\pmb{L}= 0$ would be found. It is likewise clear that more approximate solutions emerge as more and more orders are included, eventually approaching the exact set of solutions. After simple algebraic steps we verify that at order $\lambda^3$ the consistency equation \eqref{SecondSelfConsistency} reads
\begin{equation}\label{ThirdSelfConsistency}
\pmb{L} (t) = - \frac{8\lambda^3}{\pi^3} \int d^3 k \frac{\hat{k}}{k^3}|\mu_{\pmb{k}}|^2 (\pmb{L}(t) \cdot \hat{k})-\frac{\lambda}{\pi^3} \int d^3 k \left \lbrace \frac{\pmb{\epsilon}(\hat{k})\mu^*_{\pmb{k}}}{4k} + \frac{\lambda^2}{k^3} \left[ 16 (\pmb{L}(t) \cdot \hat{k})^2 - 4 |\mu_{\pmb{k}}|^2 \right]\pmb{\epsilon}(\hat{k})\mu^*_{\pmb{k}} + c.c. \right \rbrace \ .
\end{equation}
Notice that due to the definition $\mu_{\pmb{k}} = \pmb{\epsilon}(\hat{k}) \cdot \pmb{L}(t)$, each term $\mu_{\pmb{k}}$ amounts to a term linear in $\pmb{L}(t)$. Therefore the above equation is a cubic algebraic equation for the vector $\pmb{L}(t)$. The integral over the angular variables in polar coordinates, for which some details are shown in the appendix A, can be computed straight away. The result is
\begin{equation}\label{FourthSelfConsistency}
 \pmb{L}(t) = -\frac{4\lambda}{3\pi^2} \pmb{L} (t) \int dk \ k - \frac{64 \lambda^3 L^2 (t)}{15 \pi^2} \pmb{L}(t) \int \frac{dk}{k} \ .
\end{equation}
The momentum integrals are respectively ultraviolet (UV) divergent quadratically and UV and infrared (IR) divergent logarithmically.
In order to regularize the integrals, we introduce UV and IR cutoffs $Q_{UV}$ and $Q_{IR}$, so that the consistency equation can be written as
\begin{equation}\label{FifthSelfConsistency}
 \pmb{L} (t) \left \lbrace 1 + \frac{2\lambda}{3\pi^2} \left(Q^2_{UV} - Q^2_{IR} \right)+\frac{64 \lambda^3}{15 \pi^2} L^2 (t) \ln \left(\frac{Q_{UV}}{Q_{IR}} \right)\right \rbrace = 0 \ .
\end{equation}
Some comments are now in order. As anticipated, Eq. \eqref{FifthSelfConsistency} is a homogeneous cubic algebraic equation in $\pmb{L}(t)$. As a consequence the trivial solution $\pmb{L}(t) = 0$ is admitted. Similarly, due to its algebraic nature, eq. \eqref{FifthSelfConsistency} does not determine the time evolution of $\pmb{L}(t)$, only constraining its form at any given time. Thirdly, the perturbative nature of the equation is manifest, and it is understood that it represents the $O(\lambda^3)$ expansion of a trascendental exact equation, analogous to the gap equations in superconductivity \cite{BCS1,BCS2,BCS3}. We may further expect that the following terms involve UV convergent (but IR divergent) momentum integrals and that only the odd powers of $\lambda$ shall contribute. It is likewise clear that a non-trivial solution of eq. \eqref{FifthSelfConsistency} is possible only for $\lambda < 0$, i.e., in virtue of eq. \eqref{Spin-Spin Hamiltonian}, in presence of a ``ferromagnetic'' interaction that tends to align the spins (as it is the case for the torsion-induced interaction). The non trivial solution has to satisfy
\begin{equation}\label{SolutionVev}
 L^2 (t) = \frac{-\left[1 + \frac{2\lambda}{3\pi^2} \left(Q^2_{UV}-Q^2_{IR} \right) \right]}{\frac{64\lambda^3}{15 \pi^2} \ln \left(\frac{Q_{UV}}{Q_{IR}} \right)} \ ,
\end{equation}
which, considered that $\lambda < 0$, makes sense for
\begin{equation}
1+ \frac{2\lambda}{3\pi^2}\left(Q^2_{UV}-Q^2_{IR} \right) \geq 0 \ .
\end{equation}
Therefore there exists a set of non-trivial solutions $\pmb{L}(t)$ of constant square modulus given by eq. \eqref{SolutionVev}, for  coupling constant in the range
\begin{equation}\label{Coupling inequality}
 0 > \lambda \geq \frac{-3\pi^2}{2\left(Q^2_{UV}-Q^2_{IR} \right) } \ .
\end{equation}
Interestingly, the above inequality can be seen the other way around, as giving a natural bound on the cutoffs
\begin{equation}\label{Cutoff inequality}
 Q^2_{UV}-Q^2_{IR} \leq \frac{3\pi^2}{2|\lambda|} \ .
\end{equation}

\subsection{Fluctuations and validity of the mean field}

According to the Ginzburg criterion \cite{Nielsen77}, the mean field approach is sensible as long as the fluctuations around the expectation value of the order parameter can be neglected with respect to the expectation value itself. The relevant inequality reads in our case
\begin{equation}\label{GinzburgCriterion}
 |<\left(J^{5\mu}-L^{\mu}\right)\left(J^5_{\mu}-L_{\mu}\right)>| = |<J^{5\mu}J^{5}_{\mu}>-L^{\mu}L_{\mu}| \ll |L^{\mu}L_{\mu}| \ ,
\end{equation}
where the angular parentheses denote the expectation value on the condensed vacuum $\ket{0_C(t)}$ and the definition of $L^{\mu}$ has been used in the equality. The evaluation of $<J^{5\mu}J^{5}_{\mu}>$ is quite involved and it is shown in some detail in the appendix B. To lowest order the inequality \eqref{GinzburgCriterion} is, using eq. \eqref{QuadExp}
\begin{eqnarray}\label{GinzburgCriterionExp0}
 \nonumber L^2(t) \left|1- \frac{2\lambda^2}{9\pi^4}\left(Q^3_{UV}-Q^{3}_{IR} \right)\left(Q_{UV}-Q_{IR} \right) - \frac{2\lambda^2}{9\pi^4}\left(Q^2_{UV}-Q^{2}_{IR} \right)^2\right| &\ll& L^2 (t)
\end{eqnarray}
and then
\begin{eqnarray}\label{GinzburgCriterionExp}
 \left|1- \frac{2\lambda^2 }{9\pi^4}\left(Q^3_{UV}-Q^{3}_{IR} \right)\left(Q_{UV}-Q_{IR} \right) - \frac{2\lambda^2 }{9\pi^4}\left(Q^2_{UV}-Q^{2}_{IR} \right)^2\right| &\ll& 1 .
\end{eqnarray}
Neglecting $Q_{IR}$ with respect to $Q_{UV}$, the condition is satisfied when
\begin{equation}\label{FinalCutoff}
 Q^2_{UV} \simeq \frac{3\pi^2}{2|\lambda|} \ ,
\end{equation}
i.e. the mean field is more accurate the closer is the $UV$ cutoff to saturating the bound provided by eq. \eqref{Cutoff inequality}. Interestingly the mean field theory is in a certain sense self-regulating: the cutoff has to be just slightly below $\frac{3\pi^2}{2|\lambda|}$ in order for the mean field approach to be reasonable.

Notice that this has important consequences on the expectation value of eq. \eqref{SolutionVev}. Considered that the numerator of eq. \eqref{SolutionVev} is (neglecting the infrared cutoff) $-\left(1-\frac{2|\lambda|}{3\pi^2}Q^2_{UV} \right)$, the inequality \eqref{GinzburgCriterion} implies that the mean field is more accurate the smaller is the expectation value $L^2 (t)$. Still neglecting the infrared cutoff and denoting as
$
 \Delta = 1 -\frac{4\lambda^2
 }{9 \pi^4}Q^4_{UV}
$
we indeed find $L^2 (t) \propto 1 - \sqrt{1-\Delta} \simeq \frac{\Delta}{2}$, where the last approximate equality holds for $\Delta \ll 1$, i.e., in the region of validity of the mean field approach.

\subsection{Energy Gap}

From eq. \eqref{SolutionVev} we can easily compute the energy gap between the free vacuum $\ket{0}$ and the condensed vacuum $\ket{0_C(t)}$. With reference to eq. \eqref{Hamiltonian}, we have
\begin{equation}
 \bra{0} :\mathcal{H}: \ket{0} = \mathcal{E}_{0} - 2\lambda L_{\mu}(t) \bra{0} J^{5 \mu} \ket{0} + \lambda L^{\mu}(t)L_{\mu}(t) = +  \lambda L^{\mu}(t)L_{\mu}(t)
\end{equation}
where $\mathcal{E}_{0}$ is the kinetic zero point energy density, equating zero for the normal ordered (with respect to $\ket{0}$) Hamiltonian $:\mathcal{H}:$, and we have used $\bra{0} J^{5 \mu} \ket{0} = 0$. On the other hand
\begin{equation}
 \bra{0_C(t)} :\mathcal{H}: \ket{0_C(t)} = \mathcal{E}_{KIN} - 2\lambda L_{\mu}(t) \bra{0_C(t)} J^{5 \mu} \ket{0_C(t)} + \lambda L^{\mu}(t)L_{\mu}(t) = \mathcal{E}_{KIN} - \lambda L^{\mu}(t)L_{\mu}(t)
\end{equation}
where we have used the definition of $L^{\mu}(t)$, and also in this case, the Hamiltonian is normal ordered with respect to the free vacuum. At odds with the free vacuum $\ket{0}$, the condensed vacuum has a non-zero expectation value for the kinetic term $\mathcal{E}_{KIN}$ that can be computed straightforwardly using eqs. \eqref{SubVev1} and \eqref{SubVev2}. The result is, taking into account the massless limit and $L_0=0$,
\begin{equation}
 \mathcal{E}_{KIN} = \sum_r \int d^3 k \frac
 {k}{(2\pi)^3} \sin^2 \Theta_{\pmb{k},r} \simeq \frac{2\lambda^2}{(2\pi)^3} \int d^3k  \frac{|\mu_{\pmb{k}}|^2}{k} \ .
\end{equation}
Here the final equality holds to lowest order in $\lambda$ (see appendix B). The integral can be evaluated using the momentum cutoffs, yielding
\begin{equation}
 \mathcal{E}_{KIN} = \frac{\lambda^2 L^2(t)}{3 \pi^2} \left(Q^2_{UV}-Q^2_{IR} \right) \ .
\end{equation}
Integrating over a volume $V$ and taking the difference, we find the energy shift as
\begin{eqnarray}
\nonumber \Delta E &=& \int_V d^3 x \left(  \bra{0_C(t)}: \mathcal{H}: \ket{0_C(t)} -  \bra{0} :\mathcal{H}: \ket{0}\right) = V \mathcal{E}_{KIN}- 2V\lambda L^{\mu}(t)L_{\mu}(t)  \\
&=& 2V\lambda L^2\left[ 1 + \frac{\lambda}{6\pi^2}\left(Q^2_{UV}-Q^2_{IR} \right)\right] \ .
\end{eqnarray}
Considering the limitations on the ultraviolet cutoff of eq. \eqref{FinalCutoff}, neglecting the infrared cutoff and writing $\lambda = - |\lambda|$, we get
\begin{equation}
 \Delta E =  -\frac{3}{2} V |\lambda| L^2  \ .
\end{equation}
We can see that the condensed vacuum is therefore energetically favoured with respect to the free vacuum.

\section{Conclusions}

We studied the Dirac Lagrangian containing a quartic spin-spin interaction term, adopting a mean field apprach. We have shown that the quantized Hamiltonian, once diagonalized, leads to a new vacuum state, which is a condensed vacuum, energetically favored compared to the free one. We have obtained a set of self-consistency equations, discussing the limits of validity of the mean field, and computing the energy gap to leading order in the coupling.
The condensed vacuum features a non zero expectation value of the axial current.  This new source of axial current may have important implications, in particular for cosmologies that include torsion. In addition its effects may be in principle tested considering appropriate configurations in graphene \cite{Torsion1}. The analysis conducted here may be extended in several directions, eventually lifting some of the simplifying assumptions made. Both statistical considerations, regarding the phase transition to the consensed vacuum and the study of the theory at finite temperature, and cosmological considerations, based on specific torsionful models, will be the subject of future works.

\appendix

\section{Computation of the angular integrals}

The purpose of this appendix is to compute the angular part $\int_{-1}^{1}d \cos \theta \int_0^{2\pi} d \phi$ of the integrals appearing in eq. \eqref{ThirdSelfConsistency}. The helicity eigenspinors can be written as
\begin{equation}\label{HelicitySpinors}
 \xi_{+} (\hat{k}) = \begin{pmatrix}
                      \cos \frac{\theta}{2} \\
                      e^{i \phi} \sin \frac{\theta}{2}
                     \end{pmatrix} \ ; \ \ \ \
                      \xi_{-} (\hat{k}) = \begin{pmatrix}
                      -e^{-i\phi}\sin \frac{\theta}{2} \\
                      \cos \frac{\theta}{2}
                     \end{pmatrix}
\end{equation}
where it is understood that $\theta$ and $\phi$ denote the angles formed by the unit vector $\hat{k} \equiv \left(\sin \theta \cos \phi, \sin \theta \sin \phi, \cos \theta \right)$. From eq. \eqref{HelicitySpinors}, follow the vector
\begin{eqnarray}
\pmb{\epsilon} (\hat{k}) = \xi_{+}^{\dagger} (\hat{k}) \pmb{\sigma} \xi_{-}(\hat{k}) \equiv \left(\frac{1+\cos \theta - e^{-2i\phi}\left(1-\cos \theta \right)}{2} , \frac{-i - i \cos \theta -ie^{-2i\phi}\left(1 - \cos \theta \right)}{2}, -e^{-i\phi}\sin \theta \right)
\end{eqnarray}
and the scalar
\begin{eqnarray}\label{Mu}
 \nonumber \mu_{\pmb{k}} = \pmb{\epsilon}(\hat{k}) \cdot \pmb{L} (t) &=& \frac{L_x (t)}{2} \left[1+ \cos \theta - \cos 2 \phi \left(1-\cos \theta \right)  \right] - \frac{L_y(t)}{2} \sin 2 \phi (1-\cos \theta) - L_z(t) \cos \phi \sin \theta \\
 &+& i \left \lbrace \frac{L_x(t)}{2} \sin 2 \phi (1-\cos \theta) - \frac{L_y(t)}{2} \left[ 1+\cos \theta + \cos 2 \phi \left(1- \cos \theta \right)\right] + L_z(t) \sin \phi \sin \theta \right \rbrace \ .
\end{eqnarray}
The term linear in $\pmb{L}(t)$ of eq. \eqref{ThirdSelfConsistency} is proportional to
$\Re \left( \pmb{\epsilon}(\hat{k}) \mu^*_{\pmb{k}} \right)$. The $x$ component for instance reads
\begin{eqnarray*}
  \Re \left(\epsilon_x(\hat{k}) \mu_{\pmb{k}} \right) &=& \frac{L_x (t)}{4} \left \lbrace \left[ 1+ \cos \theta - \cos 2 \phi (1-\cos \theta) \right]^2 + \sin^2 2 \phi \left(1-\cos\theta \right)^2  \right \rbrace \\
 &+& \frac{L_y (t)}{4} \left \lbrace - \sin 2 \phi (1-\cos \theta) \left[1+\cos \theta - \cos 2 \phi (1-\cos \theta) \right] - \sin 2 \phi \cos 2 \phi (1-\cos \theta)^2 - \sin 2 \phi \sin^2 \theta \right \rbrace \\
&+& \frac{L_z(t)}{2} \left \lbrace - \cos \phi \sin \theta \left[1+ \cos \theta - \cos 2\phi (1-\cos\theta) \right] + \sin 2 \phi \sin \phi \sin \theta (1-\cos\theta) \right \rbrace \ .
\end{eqnarray*}
The $\phi$ integrations washes out most of the terms above, since $\int_0^{2\pi} d \phi \cos 2 \phi = \int_0^{2\pi} d \phi \sin 2 \phi = \int_0^{2\pi} d \phi \sin 2 \phi \cos 2 \phi = \int_0^{2\pi} d \phi \cos \phi \cos 2 \phi = \int_0^{2\pi} d \phi \cos \phi = \int_0^{2\pi} d \phi \sin \phi \sin 2 \phi = 0$. The only surviving terms are those proportional to $L_x(t)$, yielding
\begin{equation}
\int_{-1}^{1} d \cos \theta \int_0^{2\pi}d\phi \Re \left(\epsilon_x(\hat{k}) \mu_{\pmb{k}} \right)= \frac{8\pi}{3}L_x(t).
\end{equation}
The other components can be treated likewise, so to get
\begin{equation}\label{LinearTerm}
 \int_{-1}^{1} d \cos \theta \int_0^{2\pi}d\phi \Re \left( \pmb{\epsilon}(\hat{k}) \mu^*_{\pmb{k}} \right) = \frac{8\pi}{3} \pmb{L}(t) \ .
\end{equation}
The cubic terms are more involved, but they can be easily evaluated making use of eq. \eqref{Mu}. We quote the final results
\begin{eqnarray}\label{CubicTerm}
 \nonumber \int_{-1}^{1} d \cos \theta \int_0^{2\pi} d \phi \ \hat{k} \ |\mu_{\pmb{k}}|^2 \left(\pmb{L}(t) \cdot \hat{k} \right) &=& \frac{8 \pi L^2 (t)}{15} \pmb{L}(t) \\
 \int_{-1}^{1} d \cos \theta \int_0^{2\pi} d \phi \Re \left \lbrace \left[ 16 (\pmb{L}(t) \cdot \hat{k})^2 - 4 |\mu_{\pmb{k}}|^2 \right]\pmb{\epsilon}(\hat{k})\mu^*_{\pmb{k}} \right \rbrace &=& 0 \ .
\end{eqnarray}

\section{Vacuum fluctuations}

In this appendix we show some of the details of the computation of $<J^{5\mu}J^5_{\mu}>$. First, with reference to Eq. \eqref{ZeroComponent}, we write
\begin{eqnarray}
\nonumber  L^{\mu} &=& - \sum_{r,s} \int d^3 k \int d^3 q \ e^{i \left(\pmb{q}-\pmb{k}\right) \cdot \pmb{x}} \Bigg \lbrace A^{\mu}_{\pmb{k},\pmb{q},r,s} \bra{0_C(t)} b^{\dagger}_{\pmb{k},r} b_{\pmb{q},s} \ket{0_C (t)} +  B^{\mu}_{\pmb{k},\pmb{q},r,s} \bra{0_C(t)} b^{\dagger}_{\pmb{k},r} d^{\dagger}_{\pmb{q},s} \ket{0_C (t)} \\
 &+&  C^{\mu}_{\pmb{k},\pmb{q},r,s} \bra{0_C(t)} d_{\pmb{k},r} b_{\pmb{q},s} \ket{0_C (t)} +  D^{\mu}_{\pmb{k},\pmb{q},r,s} \bra{0_C(t)} d_{\pmb{k},r} d^{\dagger}_{\pmb{q},s} \ket{0_C (t)} \Bigg \rbrace \ .
\end{eqnarray}
The four-vectors coefficients have the form $A^{\mu}_{\pmb{k}, \pmb{q},r,s} \equiv \left(A_{\pmb{k},\pmb{q},r,s}, \pmb{A}_{\pmb{k},\pmb{q},r,s} \right)$
where the time components are explicitly given below eq. \eqref{ZeroComponent} and the spatial components are
\begin{eqnarray*}
 \pmb{A}_{\pmb{k},\pmb{q},r,s} &=& \left( \xi^{\dagger}_{r}(\hat{k}) \pmb{\sigma} \xi_s(\hat{q})\right) \left(f^*_{\pmb{k},r}f_{\pmb{q},s} + rs g^*_{\pmb{k},r}g_{\pmb{q},s}\right) \ ; \ \ \ \ \  \pmb{B}_{\pmb{k},\pmb{q},r,s} = \left( \xi^{\dagger}_{r}(\hat{k}) \pmb{\sigma} \xi_s(\hat{q})\right) \left(f^*_{\pmb{k},r}g^*_{\pmb{q},s} - rs g^*_{\pmb{k},r}f^*_{\pmb{q},s}\right) \\
 \pmb{C}_{\pmb{k},\pmb{q},r,s} &=& \left( \xi^{\dagger}_{r}(\hat{k}) \pmb{\sigma} \xi_s(\hat{q})\right) \left(g_{\pmb{k},r}f_{\pmb{q},s} - rs f_{\pmb{k},r}g_{\pmb{q},s}\right) \ ; \ \ \ \ \  \pmb{D}_{\pmb{k},\pmb{q},r,s} = \left( \xi^{\dagger}_{r}(\hat{k}) \pmb{\sigma} \xi_s(\hat{q})\right) \left(g_{\pmb{k},r}g^*_{\pmb{q},s} + rs f_{\pmb{k},r}f^*_{\pmb{q},s}\right) \ .
\end{eqnarray*}
In terms of these coefficients one has
\begin{eqnarray*}
 <J^{5\mu}J^{5}_{\mu}> \ &=& <\sum_{s_1,s_2,s_3,s_4} \int d^3 k_1 \int d^3 k_2 \int d^3 k_3 \int d^3 k_4 e^{i \left(\pmb{k_2} + \pmb{k_4} - \pmb{k_1}- \pmb{k_3} \right) \cdot \pmb{x}} \Bigg \lbrace A^{  \mu}_{\pmb{k_1},\pmb{k_2},s_1,s_2}A_{\mu ; \pmb{k_3},\pmb{k_4},s_3,s_4}b^{\dagger}_{\pmb{k_1},s_1}b_{\pmb{k_2},s_2}b^{\dagger}_{\pmb{k_3},s_3}b_{\pmb{k_4},s_4} \\
 &+& A^{  \mu}_{\pmb{k_1},\pmb{k_2},s_1,s_2}B_{\mu ; \pmb{k_3},\pmb{k_4},s_3,s_4}b^{\dagger}_{\pmb{k_1},s_1}b_{\pmb{k_2},s_2}b^{\dagger}_{\pmb{k_3},s_3}d^{\dagger}_{\pmb{k_4},s_4} + A^{  \mu}_{\pmb{k_1},\pmb{k_2},s_1,s_2}C_{\mu ; \pmb{k_3},\pmb{k_4},s_3,s_4}b^{\dagger}_{\pmb{k_1},s_1}b_{\pmb{k_2},s_2}d_{\pmb{k_3},s_3}b_{\pmb{k_4},s_4} \\
 &+& A^{  \mu}_{\pmb{k_1},\pmb{k_2},s_1,s_2}D_{\mu ; \pmb{k_3},\pmb{k_4},s_3,s_4}b^{\dagger}_{\pmb{k_1},s_1}b_{\pmb{k_2},s_2}d_{\pmb{k_3},s_3}d^{\dagger}_{\pmb{k_4},s_4} + B^{  \mu}_{\pmb{k_1},\pmb{k_2},s_1,s_2}A_{\mu ; \pmb{k_3},\pmb{k_4},s_3,s_4}b^{\dagger}_{\pmb{k_1},s_1}d^{\dagger}_{\pmb{k_2},s_2}b^{\dagger}_{\pmb{k_3},s_3}b_{\pmb{k_4},s_4} \\
 &+& B^{  \mu}_{\pmb{k_1},\pmb{k_2},s_1,s_2}B_{\mu ; \pmb{k_3},\pmb{k_4},s_3,s_4}b^{\dagger}_{\pmb{k_1},s_1}d^{\dagger}_{\pmb{k_2},s_2}b^{\dagger}_{\pmb{k_3},s_3}d^{\dagger}_{\pmb{k_4},s_4} + B^{  \mu}_{\pmb{k_1},\pmb{k_2},s_1,s_2}C_{\mu ; \pmb{k_3},\pmb{k_4},s_3,s_4}b^{\dagger}_{\pmb{k_1},s_1}d^{\dagger}_{\pmb{k_2},s_2}d_{\pmb{k_3},s_3}b_{\pmb{k_4},s_4} \\
 &+& B^{  \mu}_{\pmb{k_1},\pmb{k_2},s_1,s_2}D_{\mu ; \pmb{k_3},\pmb{k_4},s_3,s_4}b^{\dagger}_{\pmb{k_1},s_1}d^{\dagger}_{\pmb{k_2},s_2}d_{\pmb{k_3},s_3}d^{\dagger}_{\pmb{k_4},s_4} + C^{  \mu}_{\pmb{k_1},\pmb{k_2},s_1,s_2}A_{\mu ; \pmb{k_3},\pmb{k_4},s_3,s_4}d_{\pmb{k_1},s_1}b_{\pmb{k_2},s_2}b^{\dagger}_{\pmb{k_3},s_3}b_{\pmb{k_4},s_4} \\
 &+& C^{  \mu}_{\pmb{k_1},\pmb{k_2},s_1,s_2}B_{\mu ; \pmb{k_3},\pmb{k_4},s_3,s_4}d_{\pmb{k_1},s_1}b_{\pmb{k_2},s_2}b^{\dagger}_{\pmb{k_3},s_3}d^{\dagger}_{\pmb{k_4},s_4} + C^{  \mu}_{\pmb{k_1},\pmb{k_2},s_1,s_2}C_{\mu ; \pmb{k_3},\pmb{k_4},s_3,s_4}d_{\pmb{k_1},s_1}b_{\pmb{k_2},s_2}d_{\pmb{k_3},s_3}b_{\pmb{k_4},s_4} \\
 &+& C^{  \mu}_{\pmb{k_1},\pmb{k_2},s_1,s_2}D_{\mu ; \pmb{k_3},\pmb{k_4},s_3,s_4}d_{\pmb{k_1},s_1}b_{\pmb{k_2},s_2}d_{\pmb{k_3},s_3}d^{\dagger}_{\pmb{k_4},s_4} + D^{  \mu}_{\pmb{k_1},\pmb{k_2},s_1,s_2}A_{\mu ; \pmb{k_3},\pmb{k_4},s_3,s_4}d_{\pmb{k_1},s_1}d^{\dagger}_{\pmb{k_2},s_2}b^{\dagger}_{\pmb{k_3},s_3}b_{\pmb{k_4},s_4} \\
 &+& D^{  \mu}_{\pmb{k_1},\pmb{k_2},s_1,s_2}B_{\mu ; \pmb{k_3},\pmb{k_4},s_3,s_4}d_{\pmb{k_1},s_1}d^{\dagger}_{\pmb{k_2},s_2}b^{\dagger}_{\pmb{k_3},s_3}d^{\dagger}_{\pmb{k_4},s_4} + D^{  \mu}_{\pmb{k_1},\pmb{k_2},s_1,s_2}C_{\mu ; \pmb{k_3},\pmb{k_4},s_3,s_4}d_{\pmb{k_1},s_1}d^{\dagger}_{\pmb{k_2},s_2}d_{\pmb{k_3},s_3}b_{\pmb{k_4},s_4} \\
 &*& D^{  \mu}_{\pmb{k_1},\pmb{k_2},s_1,s_2}D_{\mu ; \pmb{k_3},\pmb{k_4},s_3,s_4}d_{\pmb{k_1},s_1}d^{\dagger}_{\pmb{k_2},s_2}d_{\pmb{k_3},s_3}d^{\dagger}_{\pmb{k_4},s_4} \Bigg \rbrace > \ .
\end{eqnarray*}
The expectation values can be computed by repeated application of eq. \eqref{InverseBogoliubov}. To exemplify this, consider the first term of the above equation
\begin{eqnarray*}
 && <b^{\dagger}_{\pmb{k_1},s_1}b_{\pmb{k_2},s_2}b^{\dagger}_{\pmb{k_3},s_3}b_{\pmb{k_4},s_4}> \equiv \bra{0_C(t)} b^{\dagger}_{\pmb{k_1},s_1}b_{\pmb{k_2},s_2}b^{\dagger}_{\pmb{k_3},s_3}b_{\pmb{k_4},s_4} \ket{0_C(t)} =\bra{0} \mathcal{R}(t) b^{\dagger}_{\pmb{k_1},s_1}b_{\pmb{k_2},s_2}b^{\dagger}_{\pmb{k_3},s_3}b_{\pmb{k_4},s_4} \mathcal{R}^{-1}(t) \ket{0} \\
 && = \bra{0} \mathcal{R}(t) b^{\dagger}_{\pmb{k_1},s_1}\mathcal{R}^{-1}(t) \mathcal{R}(t)b_{\pmb{k_2},s_2}\mathcal{R}^{-1}(t) \mathcal{R}(t)b^{\dagger}_{\pmb{k_3},s_3}\mathcal{R}^{-1}(t) \mathcal{R}(t)b_{\pmb{k_4},s_4} \mathcal{R}^{-1}(t) \ket{0} \\
 && = \bra{0} \left[ \cos \Theta_{\pmb{k_1},s_1}b^{\dagger}_{\pmb{k_1},s_1} + e^{-i \delta_{\pmb{k_1},s_1}}\sin \Theta_{\pmb{k_1},s_1}d_{\pmb{k_1},-s_1} \right]\left[ \cos \Theta_{\pmb{k_2},s_2}b_{\pmb{k_2},s_2} + e^{i \delta_{\pmb{k_2},s_2}}\sin \Theta_{\pmb{k_2},s_2}d^{\dagger}_{\pmb{k_2},-s_2} \right] \\
 && \times \left[ \cos \Theta_{\pmb{k_3},s_3}b^{\dagger}_{\pmb{k_3},s_3} + e^{-i \delta_{\pmb{k_3},s_3}}\sin \Theta_{\pmb{k_3},s_3}d_{\pmb{k_3},-s_3} \right]\left[ \cos \Theta_{\pmb{k_4},s_4}b_{\pmb{k_4},s_4} + e^{i \delta_{\pmb{k_4},s_4}}\sin \Theta_{\pmb{k_4},s_4}d^{\dagger}_{\pmb{k_4},-s_4} \right] \ket{0} \\
 && = e^{-i \delta_{\pmb{k_1},s_1}} e^{i \delta_{\pmb{k_4},s_4}} \sin \Theta_{\pmb{k_1},s_1} \sin \Theta_{\pmb{k_4},s_4} \bra{0} d_{\pmb{k_1},-s_1}\left[ \cos \Theta_{\pmb{k_2},s_2}b_{\pmb{k_2},s_2} + e^{i \delta_{\pmb{k_2},s_2}}\sin \Theta_{\pmb{k_2},s_2}d^{\dagger}_{\pmb{k_2},-s_2} \right] \\
 && \times \left[ \cos \Theta_{\pmb{k_3},s_3}b^{\dagger}_{\pmb{k_3},s_3} + e^{-i \delta_{\pmb{k_3},s_3}}\sin \Theta_{\pmb{k_3},s_3}d_{\pmb{k_3},-s_3} \right]d^{\dagger}_{\pmb{k_4},-s_4} \ket{0} \\
 && = e^{-i \delta_{\pmb{k_1},s_1}} e^{i \delta_{\pmb{k_4},s_4}} \sin \Theta_{\pmb{k_1},s_1} \sin \Theta_{\pmb{k_4},s_4} \bra{0} d_{\pmb{k_1},-s_1} \Bigg\lbrace \cos \Theta_{\pmb{k_2},s_2} \cos \Theta_{\pmb{k_3},s_3} b_{\pmb{k_2},s_2}b^{\dagger}_{\pmb{k_3},s_3} \\
 && + e^{i\left(\delta_{\pmb{k_2},s_2}-\delta_{\pmb{k_3},s_3} \right)} \sin \Theta_{\pmb{k_2},s_2} \sin \Theta_{\pmb{k_3},s_3} d^{\dagger}_{\pmb{k_2},-s_2}d_{\pmb{k_3},-s_3} \Bigg \rbrace d^{\dagger}_{\pmb{k_4},-s_4} \ket{0} \\
 && = e^{-i \delta_{\pmb{k_1},s_1}} e^{i \delta_{\pmb{k_4},s_4}} \sin \Theta_{\pmb{k_1},s_1} \sin \Theta_{\pmb{k_4},s_4} \cos \Theta_{\pmb{k_2},s_2} \cos \Theta_{\pmb{k_3},s_3} \delta^{3} (\pmb{k_1}-\pmb{k_4}) \delta^{3} (\pmb{k_2}-\pmb{k_3})\delta_{s_1,s_4} \delta_{s_2,s_3} \\
 && + e^{-i \delta_{\pmb{k_1},s_1}} e^{i \delta_{\pmb{k_4},s_4}}e^{i\left(\delta_{\pmb{k_2},s_2}-\delta_{\pmb{k_3},s_3} \right)} \sin \Theta_{\pmb{k_1},s_1} \sin \Theta_{\pmb{k_4},s_4} \sin \Theta_{\pmb{k_2},s_2} \sin \Theta_{\pmb{k_3},s_3} \delta^{3} (\pmb{k_1}-\pmb{k_2}) \delta^{3} (\pmb{k_3}-\pmb{k_4})\delta_{s_1,s_2} \delta_{s_3,s_4} \ .
\end{eqnarray*}
Doing so, we arrive at
\begin{eqnarray*}
 &&<J^{5\mu}J^5_{\mu}> = \sum_{s_1,s_2}\int d^3k_1 \int d^3 k_2 \Bigg \lbrace A^{\mu}_{\pmb{k_1},\pmb{k_2},s_1,s_2}A_{\mu;\pmb{k_2},\pmb{k_1},s_2,s_1} \sin^2 \Theta_{\pmb{k_1},s_1} \cos^2 \Theta_{\pmb{k_2},s_2} \\
 &&+ A^{\mu}_{\pmb{k_1},\pmb{k_1},s_1,s_1}A_{\mu;\pmb{k_2},\pmb{k_2},s_2,s_2} \sin^2 \Theta_{\pmb{k_1},s_1} \sin^2 \Theta_{\pmb{k_2},s_2} + A^{\mu}_{\pmb{k_1},\pmb{k_2},s_1,s_2}B_{\mu;\pmb{k_2},\pmb{k_1},s_2,-s_1}e^{-i\delta_{\pmb{k_1},s_1}} \sin \Theta_{\pmb{k_1},s_1} \cos \Theta_{\pmb{k_1},-s_1}\cos^2 \Theta_{\pmb{k_2},s_2} \\
 && + A^{\mu}_{\pmb{k_1},\pmb{k_1},s_1,s_1}B_{\mu;\pmb{k_2},\pmb{k_2},s_2,-s_2}e^{-i\delta_{\pmb{k_2},s_2}} \sin^2 \Theta_{\pmb{k_1},s_1} \sin \Theta_{\pmb{k_2},s_2}\cos \Theta_{\pmb{k_2},-s_2} + ... \Bigg \rbrace ,
\end{eqnarray*}
where we have omitted similar terms. To proceed with the evaluation, we expand in powers of $\lambda$, taking into account that
\begin{equation}\label{LambdaExp}
 \sin \Theta_{\pmb{k},s} \simeq \lambda \frac{|\mu_{\pmb{k}}|}{k} + 2\lambda^2 \frac{s|\mu_{\pmb{k}}|\left(\pmb{L}\cdot \hat{k} \right) }{k^2} + O(\lambda^3) \ ; \ \  \ \ \cos \Theta_{k,s} \simeq 1 - \lambda^2 \frac{|\mu_{\pmb{k}}|^2}{2k^2} + O(\lambda^3).
\end{equation}
It is straightforward, although lengthy, to check that the orders $\lambda^0$ and $\lambda^1$ yield no contribution. At order $\lambda^2$, out of all the terms appearing in the sum only four survive. We give below a detailed calculation of the first of such terms, namely
\begin{eqnarray*}
I_1 &=& \sum_{s_1,s_2} \int d^3 k_1 \int d^3 k_2  A^{\mu}_{\pmb{k_1},\pmb{k_2},s_1,s_2}A_{\mu;\pmb{k_2},\pmb{k_1},s_2,s_1} \sin^2 \Theta_{\pmb{k_1},s_1} \cos^2 \Theta_{\pmb{k_2},s_2}\\ &\simeq& \lambda^2 \sum_{s_1,s_2} \int d^3 k_1 \int d^3 k_2  A^{\mu}_{\pmb{k_1},\pmb{k_2},s_1,s_2}A_{\mu;\pmb{k_2},\pmb{k_1},s_2,s_1} \frac{|\mu_{k_1}|^2}{k_1^2} + O(\lambda^3) \ .
\end{eqnarray*}
By the definition
\begin{equation*}
 A^{\mu}_{\pmb{k_1},\pmb{k_2},s_1,s_2}A_{\mu;\pmb{k_2},\pmb{k_1},s_2,s_1} = \frac{\delta_{s_1,s_2}}{(2\pi)^6} \left(|\xi^{\dagger}_{s_1}(\hat{k}_1) \xi_{s_1} (\hat{k}_2)|^2 - |\xi^{\dagger}_{s_1}(\hat{k}_1) \pmb{\sigma}\xi_{s_1} (\hat{k}_2)|^2 \right) \ ,
\end{equation*}
so that
\begin{equation*}
 I_1 = \frac{\lambda^2}{(2\pi)^6} \int d^3 k_1 \int d^3 k_2 \frac{|\mu_{k_1}|^2}{k_1^2} \left \lbrace 8 \cos (\phi_{2}-\phi_1) \cos \frac{\theta_1}{2}  \sin \frac{\theta_1}{2}  \cos \frac{\theta_2}{2}  \sin \frac{\theta_2}{2}- 4 \cos^2 \frac{\theta_1}{2} \sin^2 \frac{\theta_2}{2} - 4 \cos^2 \frac{\theta_2}{2} \sin^2 \frac{\theta_1}{2}\right \rbrace \ .
\end{equation*}
Here the angles $\phi_j, \theta_j$ are those defining the directions $\hat{k}_j$. We perform the $d^3 k_2$ integral first in polar coordinates. The first term vanishes in the $\phi_2$ integration, while the others give
\begin{eqnarray*}
 \int d^3 k_2 \sin^2 \frac{\theta_2}{2} &=& \int_{Q_{IR}}^{Q_{UV}} dk_2 k_2^2 \int_0^{2\pi} d \phi_2 \int_{-1}^{1} d(\cos \theta_2) \left(\frac{1- \cos \theta_2 }{2}\right) = \frac{2\pi}{3} \left(Q^3_{UV}-Q^{3}_{IR} \right) \\
 \int d^3 k_2 \cos^2 \frac{\theta_2}{2} &=& \int_{Q_{IR}}^{Q_{UV}} dk_2 k_2^2 \int_0^{2\pi} d \phi_2 \int_{-1}^{1} d(\cos \theta_2) \left(\frac{1+ \cos \theta_2 }{2}\right) = \frac{2\pi}{3} \left(Q^3_{UV}-Q^{3}_{IR} \right) \ .
\end{eqnarray*}
It follows that
\begin{equation*}
 I_1 = -\frac{8\pi\lambda^2}{3 (2\pi)^6} \left(Q^3_{UV}-Q^{3}_{IR} \right) \int d^3 k \frac{|\mu_{\pmb{k}}|^2}{k^2} \ .
\end{equation*}
This last integral can be computed easily from Eq. \eqref{Mu}, so that the final result reads
\begin{equation*}
 I_1 = - \frac{\lambda^2 L^2 (t)}{9\pi^4}\left(Q^3_{UV}-Q^{3}_{IR} \right)\left(Q_{UV}-Q_{IR} \right) \ .
\end{equation*}
The other terms can be treated similarly. Overall we find
\begin{equation}\label{QuadExp}
 <J^{5\mu}J^5_{\mu}> = - \frac{2\lambda^2 L^2 (t)}{9\pi^4}\left(Q^3_{UV}-Q^{3}_{IR} \right)\left(Q_{UV}-Q_{IR} \right) - \frac{2\lambda^2 L^2 (t)}{9\pi^4}\left(Q^2_{UV}-Q^{2}_{IR} \right)^2  + O (\lambda^3) \ .
\end{equation}


\begin{thebibliography}{99}




\bibitem{QCD}
B. L. Ioffe, \textit{Physics of Atomic Nuclei} \textbf{66}, pp. 30-43 (2003).

\bibitem{QCD2}
Y. Nambu and G. Jona-Lasinio, \textit{Phys. Rev} \textbf{122}, pp. 345-358 (1961).

\bibitem{Superfluid}
A. Schmitt, \textit{Introduction to Superfluidity}, Springer (2015).

\bibitem{Superfluid2}
P. W. Anderson, \textit{Rev. Mod. Phys.} \textbf{38}, 298 (1966).

\bibitem{Superfluid3}
S. Adams and J.-B. Bru, \textit{Physica A}, \textbf{332}, pp. 60-78 (2004).

\bibitem{Superfluid4}
N. N. Bogoliubov, \textit{J. Phys. (USSR)} \textbf{11}, 23 (1947).

\bibitem{Superfluid5}
N. Angelescu, A. Verbeure and V. A. Zagrebnov, \textit{J. Phys. A: Math. Gen.} \textbf{25}, 3473 (1992).


\bibitem{BCS1}
L. N.Cooper,
\textit{Physical} Review. \textbf{104} (4): 1189–1190, (1956).


\bibitem{BCS2}
J. Bardeen, L. N. Cooper,  J. R. Schrieffer,
\textit{Physical Review} \textbf{106} (1): 162–164, (1957).

\bibitem{BCS3}
J. Bardeen, L. N. Cooper,  J. R. Schrieffer,
\textit{Physical Review} \textbf{108} (5): 1175–1204, (1957).

\bibitem{BCS4}
P. B. Allen and B. Mitrovi\'{c}, \textit{Solid State Physics} \textbf{37}, pp. 1-92 (1983).


\bibitem{Hawking}
S. W. Hawking, \textit{Commun. Math. Phys.} \textbf{43}, pp. 199-220 (1975)

\bibitem{Unruh}
W. G. Unruh, \textit{Phys. Rev. D} \textbf{14} (4), 870-892 (1976).

\bibitem{Takagi}
S. Takagi, \textit{Progress of Theoretical Physics Supplement} \textbf{88}, pp. 1-142 (1986).

\bibitem{Vanzella}
D. A. T. Vanzella and G. E. Matsas, \textit{Phys. Rev. Lett.} \textbf{87}, 151301 (2001).

\bibitem{Parker}
L. Parker, \textit{J. Phys. A: Math. Theor.} \textbf{45}, 374023 (2012).

\bibitem{Parker2}
N. Birrell and P. Davies, \textit{Quantum Fields in Curved
Space} (Cambridge Monographs on Mathematical Physics),
Cambridge University Press, Cambridge, England, 1982.

\bibitem{Casimir}
J. C. Da Silva, F. C. Khanna, A. Matos Neto and A. E. Santana, \textit{Phys. Rev. A} \textbf{66}, 052101 (2002).

\bibitem{Casimir2}
G. Plunien, B. M\"{u}ller and W. Greiner, \textit{Phys. Rep.} \textbf{134}, Issues 2-3, pp. 87-193 (1986).

\bibitem{Casimir3}
M. Bordag, G. L. Klimchitskaya, U. Mohideen and V. M. Mostepanenko, \textit{Advances in the Casimir Effect}, online edn., Oxford University Press (2009).

\bibitem{Neut1}
M. Blasone, A. Capolupo and G. Vitiello, \textit{Phys. Rev. D} \textbf{66}, 025033 (2002).


 \bibitem{CapDark1}
A. Capolupo, \textit{Adv. High En. Phys.} {\bf 2018}, 9840351 (2018).

 \bibitem{CapDark2}
A. Capolupo, \textit{Adv. High En. Phys.} {\bf 2016}, 8089142 (2016).

\bibitem{CapDark3}
A. Capolupo, S. Capozziello, G. Vitiello, \textit{Phys. Lett. A} \textbf{373}, pp. 601--610 (2009).

\bibitem{CapDark4}
A. Capolupo, S. Capozziello, G. Vitiello, \textit{Phys. Lett. A} \textbf{363}, 53 (2007).

\bibitem{CapDark5}
M. Blasone, A. Capolupo, S. Capozziello, S. Carloni, G. Vitiello, Phys. Lett. A \textbf{323}, pp. 182--189 (2004).


\bibitem{K1}
K. Fujii, C. Habe and T. Yabuki, \textit{Phys. Rev. D} \textbf{59}, 113003 (1999).

\bibitem{K2}
K. Fujii, C. Habe and T. Yabuki, \textit{Phys. Rev. D} \textbf{64}, 013011 (2001).

\bibitem{K3}
K. C. Hannabuss and D. C. Latimer, \textit{J. Phys. A: Math. Gen.} \textbf{33} 1369 (2000).

\bibitem{K4}
C. R. Ji and Y. Mishchenko, \textit{Phys. Rev. D} \textbf{65}, 096015 (2002).

\bibitem{K5}
C. R. Ji and Y. Mishchenko, \textit{Ann. Phys.} \textbf{315}, Issue 2, pp. 488-504 (2005).

\bibitem{Cap1}
M. Blasone, A. Capolupo, O. Romei, G. Vitiello, 
\textit{Phys. Rev. D} \textbf{63}, 125015, (2001).


\bibitem{Cap2}
A. Capolupo, C.R. Ji, Y. Mischenko, G. Vitiello, 
\textit{Phys. Lett. B} \textbf{594}, 135-140, (2004).


\bibitem{Axion4}
 A. Capolupo, I. De Martino, G. Lambiase and An. Stabile, \textit{Phys. Lett. B} \textbf{790}, pp. 427--435 (2019).

\bibitem{Neut2}
A. Capolupo, S. Carloni and A. Quaranta, \textit{Phys. Rev. D} \textbf{105}, 105013 (2022).

\bibitem{Neut3}
A. Capolupo and A. Quaranta, \textit{Phys. Lett. B} \textbf{839}, 137776 (2023).

\bibitem{Neut4}
A. Capolupo, A. Quaranta and R. Serao, \textit{Symmetry} \textbf{2023}, 15(4), 807 (2023).

\bibitem{Neut5}
A. Capolupo and A. Quaranta, \textit{Phys. Lett. B} \textbf{840}, 137889 (2023).

\bibitem{Neut6}
A. Capolupo, S. M. Giampaolo, G. Lambiase and A. Quaranta, \textit{Eur. Phys. J. C} \textbf{80}, 423 (2020).

\bibitem{Torsion1}

N. E. Mavromatos, P. Pais and A. Iorio, \textit{Universe} \textbf{2023}, 9 (12), 516 (2023).

\bibitem{Hehl1976}
F. W. Hehl, P. von der Heyde, G. D. Kerlick and
J. M. Nester, \textit{Rev. Mod. Phys.} \textbf{48}, 3 (1976).

\bibitem{Barker1978}
B. M. Barker and R. F. O'Connell, \textit{Gen. Rel. Grav.} \textbf{11}, 2 (1979).

\bibitem{Torsion2}
M. F. Ciappina, A. Iorio, P. Pais and A. Zampeli, \textit{Phys. Rev. D} \textbf{101}, 036021 (2020).

\bibitem{Torsion3}
I. L. Shapiro, \textit{Phys. Rep.} \textbf{357}, Issue 2, pp. 113-213 (2002).

\bibitem{Torsion4}
S. Vignolo, S. Carloni and L. Fabbri, \textit{Phys. Rev. D} \textbf{91}, 043528 (2015).

\bibitem{Torsion5}
L. Fabbri and S. Vignolo, textit{Class. Quantum Grav.} \textbf{28}, 125002 (2011). 

\bibitem{Torsion6}
S. Capozziello, M. De Laurentis, L. Fabbri and S. Vignolo, \textit{Eur. Phys. J. C} \textbf{72}, 1908 (2012).


\bibitem{Propagating1}
D. Benisty, E. I. Gundelman, A. Van de Venn, D. Vasak, J. Struckheimer and H. Stoecker, \textit{Eur. Phys. J. C} \textbf{82}, 264 (2022).

\bibitem{Propagating2}
S. M. Carrol and G. B. Field, \textit{Phys. Rev. D} \textbf{50}, 3867 (1994).

\bibitem{Propagating3}
Y.-F. Cai, S. Capozziello, M. De Laurentis and E. N. Saridakis, \textit{Rep. Prog. Phys.} \textbf{79}, 106901 (2016).


\bibitem{SUGRA}
P. Van Nieuwenhuizen, \textit{Phys. Rept.} \textbf{68}, 4, pp. 189-398 (1981).

\bibitem{Heis1}
I. Affleck, \textit{J. Phys.: Condens. Matter} \textbf{1}, 3047 (1989)

\bibitem{Heis2}
B. Simon, \textit{The Statistical Mechanics of Lattice Gases}, Princeton University Press, Princeton, N.J. (1993).











\bibitem{Nielsen77}
J. Als-Nielsen and R. J. Birgeneau, \textit{Am. J. Phys.} \textbf{45}, 554-560 (1977).

\bibitem{Umezawa1}
H. Umezawa, H. Matsumoto, Masashi Tachiki, \textit{Thermo field dynamics and condensed states}, North-Holland Publishing Company, 1982, ISBN 978-0444863614

\bibitem{Umezawa2}
H. Umezawa, \textit{Advanced Field Theory: Micro, Macro, and Thermal Physics}, American Institute of Physics, 1993, ISBN 978-1563960819

\bibitem{Campbell}
 J. E. Campbell, Proceedings of the London Mathematical Society \textbf{28}, 381–390 (1897);
 Proceedings of the London Mathematical Society \textbf{29}, 14–32 (1898).
  \\
  H. F. Baker, Proceedings of the London Mathematical Society (1) \textbf{34}, 347–360 (1902);
Proceedings of the London Mathematical Society  (1) \textbf{35}, (1903) 333–374;
Proceedings of the London Mathematical Society (Ser 2) \textbf{3}, 24–47 (1905).
 \\
 F. Hausdorff, "Die symbolische Exponentialformel in der Gruppentheorie", Ber Verh Saechs Akad Wiss Leipzig \textbf{58}, 19–48 (1906).
 \\
A. Perelomov, \textit{Generalized coherent states and their applications}, Springer, Berlin 1986.



\end{thebibliography}
\end{document}